\newcommand{\T}{{\cal{T}}}
 \documentstyle[12pt]{article}
\newcommand{\be}{\begin{equation}}
\newcommand{\ee}{\end{equation}}

\begin{document}
\begin{center}{\bf \Large Exactly unsolved problems \\
of interacting 1D fermions}\end{center}
\begin{center} {\it Sergei Skorik}
\footnote{
The author was supported
 by Kreitman and by Kaufmann fellowships, and partially funded
by the Israel Science Foundation, grant No 3/96-1. These notes were
inspired by the discussions with Natan Andrei, Yuval Geffen and Yigal Meir.}
             \end{center}
\begin{center} {\small \it 
 Department of Physics, Ben-Gurion University, Beer-Sheva 84105,
Israel, and \\
 Department of Physics, Weizmann Institute for Science, 
Rehovot 76100, Israel}
\end{center}

\vspace{0.5cm}

\noindent
Applications of the integrable system 
techniques to the non-equilibrium transport
problems are discussed. We describe one-dimensional 
electrons tunneling through a point-like 
defect either by the s-d exchange (Kondo)
mechanism, or via the resonanse level (Anderson) mechanism. These models are 
potential candidates 
to be solved exactly in the presence of arbitrary external bias.
We draw attention also 
to several mesoscopical systems which can be tackled by the 
massless form-factor approach, as perturbations of integrable models. The basic 
unperturbed model is the massless 
sine-Gordon model with the interaction  (cosine) 
term restricted to one point, which is integrable. It is being perturbed by the 
second interaction term, 
which destroys integrability. Quasi-exact results can be 
obtained by making use of the basis of 
massless quasiparticles of the sine-Gordon 
model.

\vspace{1cm}

\begin{center} {\small INTRODUCTION} \end{center}

\vspace{3mm}

\noindent Bethe ansatz technique
(BA) is a powerfull tool for solving strongly interacting systems. Among
its most interesting applications are the exact solutions of Kondo and
Anderson models of impurities in the metals, sin-Gordon/Thirring
model, etc. The central issue of the Bethe ansatz is a phenomenon
of factorized scattering which allows for solving the theory completely
starting from the two-particle scattering matrices as the only input data. 
Factorized scattering approach turned out to be more fruitful 
than thought earlier in the traditional  frame of BA.
Withing the standard scope of BA, as we define it,  lies
solving  the bare Hamiltonian for its excitation spectrum, the distribution
functions and, eventually, the partition function, which contains
the full information on the equilibrium properties. 
Factorized scattering concept, 
however, led further to the development of new methods,
 providing us with additional important physical
quantities: the series expansion of the correlation functions that
follows from the form-factor approach and the
 finite-size ground state energy of Zamolodchikov's thermodynamic BA.  
The attitude which is involved here (as opposed to the traditional BA)
is similar in spirit to the Landau
theory of Fermi liquids: the fundamental objects which characterize
the behavior of the system are the dressed quasiparticles,
 rather than the bare particles
of the free problem. Free particles, after the interactions have been turned
on, will acquire the screening cloud of virtual pairs that will follow it,
resulting in the formation of new objects -- quasiparticles that substitute
the free electrons in the non-interacting system.  
The spectrum, density and interactions of quasiparticles
completely determine the physical properties of the system. In practice it is
hardly possible to determine the properties of quasiparticles exactly,
starting from the non-interacting system and switching on interactions,
except for the special integrable cases, when the Bethe ansatz is applicable.
Integrable quasiparticles are, however, very special as far as 
their interactions are concerned: they are stable, meaning that the conservation
laws force the momentum conservation of each individual quasiparticle,
and their scattering is factorized into the two-particle processes
(in some more exotic integrable 
models there can in principle exist quasiparticles
with imaginary energy - finite lifetime, so-called monstrons,
 but we do not consider them here).
Adopting integrable quasiparticle basis, one can do the best of it
in describing also non-integrable models as perturbations of the integrable
ones. This direction was 
considered in the works of Delfino, Mussardo and Simonetti
[Nucl.Phys. B473 (1996) 469]. The quasiparticles approach turned out also to be
effective in the exact description of non-equilibrium properties of integrable
systems as shown in the works of N.Andrei [Phys.Lett. A87 (1982) 299], 
Fendley, Ludwig and Saleur [Phys.Rev. B52 (1995) 8934], Peres, Sacramento
and Carmelo [cond-mat/9709144]. In the present notes we 
are trying to pursue further this approach and 
sketch a few possible applications
of integrable quasiparticle techniques to the tunneling transport problems,
limiting ourselves to the technically simplest cases that naturally
lead to more realistic and complicated models.

\vspace{1cm}

\begin{center} {\small I. EXACT NON-LINEAR TUNNELING CURRENT} \end{center}

\vspace{3mm}

\noindent In this section we describe two 1D models of electrons 
tunneling 
 through a point-like defect either by the s-d exchange (Kondo)
mechanism, or via the resonanse level (Anderson) mechanism. These  models
are potential candidates to be solved exactly in the presence
of arbitrary external bias by means of the Bethe-ansatz technique.

\vspace{0.5cm}

 1. {\it s-d exchange model of zero-bias tunneling anomalies.}

\vspace{5mm}

\noindent The  model we focuse on
dates back to the original work of J.Appelbaum [Phys.Rev.Lett. 17 (1966) 91].
Imagine two pieces of metal with different
chemical potentials, separated by an insulating material. 
Tunneling of electrons is made possible at one special point, 
 as if the two metals were connected
to each other adiabatically
by a ``weak link.'' Suppose that,  by integrating out transversal modes,
 one can reduce the
problem to effectively one-dimensional, with a point-like impurity at 
$z=0$. The precise nature
of the transversal modes is not essencial as far as only the low-energy physics
is concerned and the transitions between different modes
can be neglected. Consider first the right
hand side metal. In the low-energy limit one can linearize the electron
spectrum near the Fermi energy, leading to the Hamiltonian
$H=\sum_\epsilon \epsilon a_{R\epsilon}^+a_{R\epsilon} + \epsilon
a_{L-\epsilon}^+a_{L-\epsilon}$. It is important that the states responsible
for the low-energy physics can be labeled by a single parameter, $\epsilon$.
It is convenient to introduce 
\be \Psi=\Psi_L + \Psi_R, \qquad z>0\ee
where $\Psi_{L,R}$ are the one-dimensional left and right moving electron
fields, build by the wave-functions of effective 1D problem: 
$\Psi_{L,R}=\pm\sum_\epsilon e^{\pm i\epsilon z} e^{i\epsilon t}a_\epsilon$. 
These fields obey the usual anti-commutation relations.
With the
boundary condition imposed,
\be \Psi(0)=\Psi_L(0)+\Psi_R(0)=0,\ee
 situation corresponding to the full reflection. One can map one of the fields,
say, right-moving, to the negative semi-axis preserving the boundary
condition at zero  as follows
\be \tilde{\Psi}_L(-z)=-\Psi_R(z)\ee
\be\Psi_L(0)=\tilde{\Psi}_L(0)\ee
The same ``unfolding'' procedure can be applied to the left piece
of metal to map the problem eventually to the full axis. We end up
with two species of free Dirac electrons on the line, to which we assign
the flavor index $m=1,2$. The Hamiltonian describing this problem is:
\be
H_0= iv_F\sum_{\sigma,m=1,2}\int_{-\infty}^\infty\Psi^+_{\sigma m}
\partial_z\Psi_{\sigma m} dz + \sum_{\sigma,m=1,2}\mu_m\int_{-\infty}^
\infty \Psi^+_{\sigma m}\Psi_{\sigma m} \label{HAMI}
\ee
where we introduced the bias term which accounts for the fact that
the left and right metals are at different chemical potentials
$\mu_{1,2}$. We set $\mu_1=-\mu_2=\mu$ and rewrite the last term as
\be
Y_0=\mu\sum_\sigma\int_{-\infty}^\infty(\Psi_{1,\sigma}^+\Psi_{1,\sigma}
-\Psi^+_{2,\sigma}\Psi_{2,\sigma}) \label{BIAS}
\ee
The Hamiltonian (\ref{HAMI}) possesses $SU(2)^{flavor}$ and $SU(2)^{spin}$
symmetries, as well as $U(1)^{charge}$ symmetry, 
and the total number of type 1 and type 2 particles 
is conserved separately. Now we introduce a weak tunneling between
the metals via an impurity spin $S=1/2$ at the point $z=0$:
\be
H_{int}=\sum_{\lambda=1}^3\sum_{ab} 
J^{ab}_\lambda\Psi^+_{\sigma a}(0)\sigma^\lambda_{\sigma\sigma'}
\Psi_{\sigma' b}(0)S^\lambda \label{INT}
\ee
where $\sigma^\lambda$ are Pauli matrices and $J_\lambda^{ab}$ are
coupling constants. We make a natural assumption that $J_x^{ab}=J_y^{ab}$
and $J^{12}=J^{21}, \quad J^{11}=J^{22}$.
The full Hamiltonian $H_0+H_{int}$ resembles the two-channel Kondo model,
with however additional intra-channel interaction.
The interaction term $J^{12}$ breaks the $SU(2)^{flavor}$ symmetry down to
its $U(1)^{flavor}$ subgroup. As a resut,
the number of type 1 and type 2 particles is no longer
conserved separately; however, the total number of type 1 and type 2
particles is still a good quantum number:
\be {\cal{Q}}^{charge}=
\int \Psi_1^+\Psi_1+\Psi_2^+\Psi_2 \label{Q}
\ee
 The ${\cal{S}}^{flavor}_x$
charge corresponding to $U(1)^{flavor}$ subgroup also remains conserved:
\be {\cal{S}}_x^{flavor}=\sum_{ab}\int\Psi_a^+\tau_x^{ab}\Psi_b=
\int \Psi_1^+\Psi_2+\Psi_2^+\Psi_1 \label{SX}
\ee
Here $\tau_x$ is the first Pauli matrix acting in the flavor space,
and the summation over the spin index is implicitly assumed.

We are not going  to discuss the experimental relevance of this particular
model, i.e. what physical systems it actually describes. We believe that
it is one of the simplest and yet non-trivial theoretical models,
and therefore it is instructive to learn as much as possible from it.
We shall mention, however, that more complicated and realistic models,
such as the one introduced by Vladar and Zawadowski [Phys.Rev. B28 (1983) 1564],
are its direct generalizations as far as the methodology is concerned. 
The reader interested in the relevant experimental advances
can see the papers of Ralph and Buhrman [Phys.Rev.Lett. 69 (1992) 2118,
Phys.Rev. B51 (1995) 3554]. Some exact theoretical results
in this field, based on the conformal field
theory and the Hershfield's formalism, include the exact scaling
functions found by Delft, Ludwig and Ambegaokar [cond-mat/9702049, 
submitted to Annals of Physics]. It is worth mentioning that the one-dimensional
approximation that we adopted may not describe correctly certain
microscopic quantities, such as the local density of states in the
realistic point contacts, as was noticed by Ulreich and Zwerger 
[cond-mat/9710174,
submitted to Europhys.Lett.]. However, it does work extremely well
for the linear conductance at zero temperature [Phys.Rev. B50 (1994) 17320].

One can bosonize the fermions according to the standard procedure and
express the Hamiltonian in terms of four left-moving boson fields:
\be\Psi^+_{\sigma m}=:e^{i\varphi_{\sigma m}}:\ee
Following Emery and Kivelson [Phys.Rev.B47 (1992) 10812],
introduce linear combinations of the bosons as follows:
\be\phi_c={1\over2}(\varphi_{\uparrow 1}+\varphi_{\downarrow 1}+
\varphi_{\uparrow 2}+\varphi_{\downarrow 2})\ee
\be\phi_s={1\over 2}(\varphi_{\uparrow 1}-\varphi_{\downarrow 1}
+\varphi_{\uparrow 2} -\varphi_{\downarrow 2})\ee
\be\phi_f={1\over 2}(\varphi_{\uparrow 1}+\varphi_{\downarrow 1}
-\varphi_{\uparrow 2}-\varphi_{\downarrow 2})\ee
\be\phi_{sf}={1\over 2}(\varphi_{\uparrow 1}-\varphi_{\downarrow 1}
-\varphi_{\uparrow 2}+\varphi_{\downarrow 2})\ee
The field $\phi_c$ represents the total charge degree of freedom
in two channels and decouples from the Hamiltonian. The resulting
Hamiltonian depends on the remaining three fields as follows:
\be H_0={v_F\over 2\pi}\sum_{m=s,f,sf}\int(\partial_z\phi_m)^2+{\mu\over\pi}
\int\partial_z\phi_f\ee
\be
H_{int}=4[J_x^{12}\cos\phi_f(0)+J_x^{11}\cos\phi_{sf}(0)]\cdot
[\cos\phi_s(0)S^x+\sin\phi_s(0)S^y] \label{INNT} \ee
\be -4J_z^{12}\sin\phi_f(0)\sin\phi_{sf}(0)S^z+
{J_z^{11}\over \pi}\partial_z\phi_s(0)S^z \nonumber \ee
Interaction term (\ref{INNT}) resembles the one-channel Kondo
model where the coupling constant $J_x$ became a dynamical degree of freedom,
$J_x\to [J_x^{12}\cos\phi_f(0)+J_x^{11}\cos\phi_{sf}(0)]$.
There exists a special point in the space of couplings
($J_z^{11}=2\pi v_F$ and $J_z^{12}=0$), 
an analogue of the Toulouse point in the Kondo model, 
where the full Hamiltonian $H=H_0+H_{int}$
becomes quadratic in terms of new fermion operators and can be solved exactly. 
 This was noticed by Schiller and 
Hershfield [Phys.Rev.B51 (1995) 12896], who found the non-equilibrium
current at this ``Toulouse point.'' 

In the absense of bias, the full interacting Hamiltonian 
(\ref{HAMI})-(\ref{INT}) is integrable. To see this, let us make a canonical
transformation to another basis of fermions:
\be
a_\sigma={1\over\sqrt{2}}(\Psi_{1\sigma}+\Psi_{2\sigma}) \label{NEWI}
\ee
\be
b_\sigma={1\over\sqrt{2}}(-\Psi_{1\sigma}+\Psi_{2\sigma}) \label{NEWII}
\ee
In this new basis the matrix of couplings $J^{ab}$ becomes diagonal
with the eigenvalues $J^\pm=J^{11}\pm J^{12}$. The resulting Hamiltonian
is the channel-anisotropic two-channel Kondo model, which was solved by means
 of Bethe ansatz by N.Andrei and A.Jerez [Phys.Rev.Lett.74 (1995)
4507]:
\be
H=iv_F\sum_\sigma\int(a_\sigma^+\partial_za_\sigma +
b_\sigma^+\partial_zb_\sigma) + \sum_\lambda [J^+_\lambda a^+_\sigma(0)\sigma_
{\sigma\sigma'}^\lambda a_{\sigma'}(0)+J^-_\lambda b^+_\sigma(0)\sigma_
{\sigma\sigma'}^\lambda b_{\sigma'}(0)]S^\lambda \label{HAMII}
\ee
\be Y_0=\mu\sum_\sigma\int a^+_\sigma b_\sigma + b^+_\sigma a_\sigma\ee
The advantage  of this new basis is that the number of a-particles
as well as b-particles is a conserved quantity, and the Bethe wave functions
can be constructed in the sector of fixed quantum numbers $N_a$ and $N_b$.
These conserved charges correspond to the total 
particle number ${\cal{Q}}^{charge}$ and ${\cal{S}}^{flavor}_x$ 
in the old basis of $\Psi$
($U(1)^{charge}$ and $U(1)^{flavor}$ symmetries).
Operator $Y_0$, being a generator of the $SU(2)^{flavor}$ symmetry,
 does not commute with the Hamiltonian, except
for the channel-isotropic point $J^+=J^-$, corresponding to $J^{12}=0$,
where the $SU(2)^{flavor}$ symmetry is restored (note that in general
$J^+\geq J^-$). The non-conservation of $SU(2)^{flavor}$ charge,
in particular of its component $Y_0$, is responsible for the transport
properties we are interested in. 

Following Andrei and Jerez, let us study one and two-particle sectors
of the first-quantized analogue of Hamiltonian (\ref{HAMII}). Naively,
the scattering matrix on impurity would be
$$R_a^a=\exp(iJ^+\vec{\sigma}\vec{S}), \quad 
R_b^b=\exp(iJ^-\vec{\sigma}\vec{S}), \quad R_a^b=0
$$
where the scattering is obviously diagonal in the flavor space.
Respectivly, the scattering matrix on impurity dictates us what
the scattering matrix in the bulk would be, since the basis of the
wave-functions in the bulk is not unique and must be chosen in the consistent
way with the impurity. Let us clarify this point. Consider one of the
possible representations of the
 two-particle
wave function for free 1D fermions with the linear spectrum:
$$\psi(x_1,x_2)=\theta(x_1-x_2)(A_{\sigma_1\sigma_2}e^{ik_1x_1+ik_2x_2}
+B_{\sigma_1\sigma_2}e^{ik_1x_2+ik_2x_1})$$
$$+\theta(x_2-x_1)((SA)_{\sigma_1\sigma_2}e^{ik_1x_1+ik_2x_2}+
(SB)_{\sigma_1\sigma_2}e^{ik_1x_2+ik_2x_1})$$
Here $A$ and $B$ are spinors, $S$ is for the time being arbitrary matrix acting
on the spinors, and $\theta(x)$ is a step function. It is easy to check
that $\psi(x_1,x_2)$ is a solution of Schrodinger equation
$$(-i\partial_{x_1}-i\partial_{x_2})\psi=E\psi$$
with $E=k_1+k_2$. In addition, one must require $\psi(x_1,x_2)$ to be 
antisymmetric with respect to the simultaneous 
permutation of spin and coordinate indexes.
This leads to the relation between $A$ and $B$, 
$$B=-PSA$$
and the constraint for $S$:
\be
S_{12}S_{21}=1 \label{CONSTR}
\ee
where $P$ is the permutation operator acting on spinors and
$S_{21}=PS_{12}P$. The condition (\ref{CONSTR}) leaves some freedom for us
to choose the matrix $S$ (for example, $S=I$ and $S=P$ are valid solutions
to (\ref{CONSTR})).
The consistency between impirity and bulk scattering
 is represented by the boundary
Yang-Baxter equation,
$$S_{12}R_2R_1=R_1R_2S_{12}$$
which imposes a further constraint on $S$, or in other words selects
bulk scattering in a consistent way with the impurity.
With the impurity scattering matrix given above, the Yang-Baxter equation 
would be solved by the bulk S-matrix 
$$ S_{12}=P^{spin}P^{flavor}$$
where $P$ are exchange operators acting in spin or flavor spaces
of two particles. As noticed by Wiegmann and Tsvelik, as well as by Andrei
and Destri, such a choice of scattering matrices leads to wrong results.
The Hamiltonian must be regularized appropriately from the very beginning,
which modifies S-matrices in such a way that they become momentum-dependent
(it was not possible before regularization because J is a dimensionless
coupling and there is no scale in the problem; after the regularization
the model acquires a scale which enters S-matrices in the dimensionless
combination with momentum). Andrei and Destri [Phys.Rev.Lett 52
(1984) 364] 
suggested an elegant regularization compatible with the integrability:
one introduces the second-order 
derivative term coupled to a mass scale $\Lambda$
 in the first-quantized Hamiltonian, $\Lambda^{-1}\partial^2$, thus 
``delinearizing'' the spectrum.
After such a regularization the bare scattering matrices become:
\be R_a^a(k)={k/\Lambda +1 -iJ^+(\vec{\sigma}\vec{S}+1/2)\over
 k/\Lambda +1 -iJ^+}\ee
\be R_b^b(k)={k/\Lambda +1 -iJ^-(\vec{\sigma}\vec{S}+1/2)\over
 k/\Lambda +1 -iJ^-}\ee
(impurity scattering is still flavor-diagonal)
\be S_{12}^{spin}(k_1,k_2)={(k_1/\Lambda+1)/J^\pm-(k_2/\Lambda+1)/J^\pm-
iP^{spin}
\over (k_1/\Lambda+1)/J^\pm-(k_2/\Lambda+1)/J^\pm-i}\ee
where $J^+ (J^-)$ is chosen depending on the flavor of the first and second
particles,
\begin{eqnarray}
S_{12}^{flavor}(k_1-k_2)
&=&{i\over 2}{\sin\nu\over\sinh[\kappa(k_1-k_2)/\Lambda+i\nu]}
[\tau_x\otimes\tau_x + \tau_y\otimes\tau_y]
+{1\over 2}[1+\tau_z\otimes\tau_z] \nonumber \\
&+&{1\over 2}{\sinh[\kappa(k_1-k_2)/\Lambda]
\over\sinh[\kappa(k_1-k_2)/\Lambda+i\nu]}[1-\tau_z\otimes\tau_z]
\end{eqnarray}
Here $\tau$ are the Pauli matrices acting on the 
flavor space, $\kappa$ and $\nu$
are functions of $J^\pm$. In the scaling limit we have $\nu/\kappa=J^-$.
The complete bulk scattering matrix is a tensor product of spin and
flavor terms, $S_{12}=S_{12}^{spin}\otimes S_{12}^{flavor}$.
The $ S_{12}^{spin}$ term reflects the $SU(2)^{spin}$ invariance, while
the flavor component $S_{12}^{flavor}$ reflects the breaking of the
$SU(2)^{flavor}$ symmetry to $U(1)^{flavor}$ subgroup in the presence
of anisotropy. 

What is the form of the above S-matrices in the original basis of
$\Psi$-fields? Since our transformation 
(\ref{NEWI})-(\ref{NEWII}) mixes up the flavors (and not the
spins), we will list here only two of the above terms mapped to the original
basis:
\be\tilde{R}_1^1=\tilde{R}_2^2={1\over 2}(R_a^a+R_b^b)   \ee
\be\tilde{R}_1^2=\tilde{R}_2^1= {1\over 2}(R_a^a-R_b^b)  \ee
The impurity scattering is obviously non-diagonal, allowing the processes
of flavor violation $1\to 2$, which accounts for the non-trivial transport
through impurity from the left to the right metal. Further, the bulk 
scattering becomes
\begin{eqnarray}
\tilde{S}_{12}^{flavor}&=&
{i\over 2}{\sin\nu\over\sinh[\kappa(k_1-k_2)/\Lambda+i\nu]}
[\tau_z\otimes\tau_z + \tau_y\otimes\tau_y]
+{1\over 2}[1+\tau_x\otimes\tau_x]
\nonumber \\
&+&{1\over 2}{\sinh[\kappa(k_1-k_2)/\Lambda]
\over\sinh[\kappa(k_1-k_2)/\Lambda+i\nu]}[1-\tau_x\otimes\tau_x]
\end{eqnarray}
The non-conservation of flavor is also explicitly present
 in this term: the in-state
$|1\rangle\otimes|1\rangle$ will be mapped to the supperposition of 
$|1\rangle\otimes|1\rangle$ and $|2\rangle\otimes|2\rangle$. So,
the scattering of particles of first kind into particles of the second
kind occurs not only on the impurity, but also in the bulk. 

\vspace{1cm}

2. {\it Non-equilibrium transport through a resonance level.}

\vspace{5mm}

\noindent Anderson proposed a more generic model than Kondo-type model 
described in the previous section. 
 All the arguments 
 of the previous section up to formula (\ref{INT}) should be recalled
as they are unchanged here. Following
Anderson, one assumes the existence of localized states (either impurity
or surface states) on the interface between two metals. Such states,
or for our purposes just one state  is introduced
formally as $d_\sigma^+|0\rangle$, and the
coupling to bulk electrons takes a form (instead of equation (\ref{INT}))
\be
H_{int}=V\sum_{\sigma,m}[\Psi^+_{\sigma m}(0)d_\sigma+d^+_\sigma\Psi_{
\sigma m}(0)] \label{INTI}
\ee
Besides, one adds to Hamiltonian the part which describes the localized level:
\be
H_d=\epsilon_d\sum_{\sigma=\pm} d^+_\sigma d_\sigma + 
U\sum_{\sigma=\pm} d^+_\sigma
d_\sigma d^+_{-\sigma}d_{-\sigma} \label{INTD}
\ee

The stationary one-particle eigenstate of $H_0+H_{int}+H_d$ is a superposition
of three terms,
\be
|\psi_\sigma\rangle=\int dx [g_1(x)\Psi_{1\sigma}^+(x)+g_2(x)\Psi_{2\sigma}
^+(x)] |0\rangle + ed^+_\sigma |0\rangle 
\ee
(the particle can be either on the localized level or in the bulk in one
of two flavors with spin preserved).
From it we infer the impurity scattering matrix
\be R_1^1=R_2^2=R_1^2=R_2^1=e^{i\Delta(E)} \ee
\be \Delta(E)=-2\tan^{-1}{V^2\over E-\epsilon_d} \ee

The Anderson model has $U(1)^{flavor}$ symmetry, like the Kondo
model of the previous section. Notice, however, that
the flavor $SU(2)$-symmetric two-channel Anderson model can be used to get
the above  model (\ref{INTI})-(\ref{INTD}). The $SU(2)$-symmetric
Anderson model differs by the presence of two localized levels,
$d_{1\sigma}^+|0\rangle$ and  $d_{2\sigma}^+|0\rangle$, 
each one coupled to its channel,
\be
H_{int}=V\sum_{\sigma, m}[\Psi^+_{\sigma m}(0)d_{\sigma m}+
d_{\sigma m}^+\Psi_{\sigma m}(0)]
\ee 
\be
H_d=\epsilon_d\sum_{m,d}d^+_{m\sigma}d_{m\sigma}-{U\over 2}\sum_{mm'\sigma
\sigma'} d^+_{m\sigma}d^+_{m'\sigma'}d_{m'\sigma}d_{m\sigma'}
\ee
In the $SU(2)$-symmetric model the number of particles of each kind
is conserved separately (particles in the bulk and on the localized
level must be added together to obtain a conserved charge), and
moreover, the $SU(2)$ symmetry in the flavor space is present. To break
this symmetry down to $U(1)$ and to destroy the conservation of each
kind of particles separately, thus reproducing the original Anderson model
(\ref{INTI})-(\ref{INTD}), we introduce an additional term, describing
hopping between localized levels:
\be
H_{12}=-t\sum_\sigma(d_{1\sigma}^+d_{2\sigma} + d_{2\sigma}^+d_{1\sigma})
\ee
Then one must take the limit $t\to\infty$ to reproduce the model of
interest (intuitevely, this limit means identifying $d_1$ with $d_2$ and
calling both of them by $d$).\footnote{When hopping amplitude $t$ is very
large, it is worth to work in the basis of mixed states
$e_1=d_1-d_2, \quad e_2=d_1+d_2$, where the hopping and $\epsilon_d$
-terms become diagonal: $(\epsilon_d+t)e_1^+e_1 +(\epsilon_d-t)e_2^+e_2$.
In the limit $t\to\infty$ one of the mixed localized levels $(e_1)$ 
becomes so high in energy that the transitions to this level are unlikely
to take place and this level decouples. Then, after rescaling 
$\epsilon_d$ one can identify $e_2=d$ (the spin index is assumed everywhere).}

The same discussion about conserved charges as in the previous
section applies to Anderson model. In particular, after the same canonical
transformation of fields (\ref{NEWI})-(\ref{NEWII}) we obtain
the Anderson model with one kind of particles coupled to impurity,
while another kind of particles being free. These two kinds of particles
are, however, coupled to each other by the bias term:
\be H_a=iv_F\sum_\sigma\int a_\sigma^+\partial_za_\sigma +
V\sum_\sigma[a^+_\sigma(0)d_\sigma+d_\sigma^+a_\sigma(0)]
+\epsilon_d\sum_\sigma d^+_\sigma d_\sigma + U\sum_\sigma d^+_\sigma
d_\sigma d^+_{-\sigma}d_{-\sigma} \ee
\be H_b=iv_F\sum_\sigma\int b_\sigma^+\partial_zb_\sigma \ee
\be Y_0=\mu\sum_\sigma\int a^+_\sigma b_\sigma + b^+_\sigma a_\sigma\ee

\vspace{1cm}

3. {\it Open questions.}

\vspace{5mm}

\noindent Computing the non-equilibrium current 
implies an evaluation of the average
\be
I=e{d\over dt}\langle Y_0 \rangle =e\mu{d\over dt}\sum_\sigma\int\langle
a_\sigma^+b_\sigma+b^+_\sigma a_\sigma \rangle \label{SK}
\ee
on the Bethe states. 
Fendley, Ludwig and Saleur [Phys.Rev. B52 (1995) 8934] suggested
that at least for certain integrable models (in their case
quantum Hall bar with a constriction) the following procedure can be
used. First, based on the Bethe ansatz solution in the absense
of bias, one identifies bulk excitations (physical quasiparticles) that
carry a non-trivial charge/flavor. Further, using standard techniques,
one calculates scattering matrices in the bulk and on the impurity for
the physical quasiparticles and, in the absense of impurity but in the
presence of bias, densities of states. Then, based on the probabilistic
arguments, one employes the Bolzmann rate equation to obtain the current:
\be
I\sim\int d\epsilon [n_1(\epsilon,\mu)-n_2(\epsilon,\mu)] |S_{12}(\epsilon)|^2
\label{BEQ}
\ee
Note that this procedure has not been shown rigorously
to coincide with (\ref{SK}), 
and remains as a conjecture [cond-mat/9708163, 9710205].
Working in the original $\Psi$-basis, we face the problem of non-diagonal
bulk scattering  -- the scattering with the flavor violation $1\to 2$
 happens everywhere, and not only on the
impurity, as it was in the case of the
Quantum Hall problem with a constriction. The flavor z-projection
$Y_0$ is not diagonal on the Bethe ansatz states.
This leads to some technical
difficulties, as well as to the question of the applicability of (\ref{BEQ})
for such situations. 
Equivalently, the difficulty can be understood 
in the representation (\ref{HAMII}) of a and b-particles and Eq (\ref{SK}).
Operator $Y_0$ 
(x-component of the flavor in this representation) strongly disturbs
the Bethe state and does not allow us to work only with a few significant
low-lying excitations. The state obtained after applying $Y_0$ to the Bethe
vacuum state involves infinite amount of excitations in the thermodynamic 
limit.
The situation is rather similar to the XXZ chain
and the $S^x$ operator, which perturbs spins on all sites simultaneously,
while the Bethe states are constructed as local spin-flips with respect to some
reference state. Thus, one says that $S^x$ excites an infinite
amount of elementary particles. 

To summurize, the questions to be answered are:
 what are the physical excitations
in the Bethe ansatz solution of (\ref{HAMI})-(\ref{INT}) and their scattering
matrices?  What are the matrix elements 
of the fields between the excited states (form-factors for the 
two-channel Kondo model)? Can one apply Eq (\ref{BEQ}) to the model
(\ref{HAMI})-(\ref{INT})?

\vspace{1cm}

\begin{center}  {\small II. COULOMB BLOCKADE AND THE INTERFERENCE} \end{center} 

\vspace{3mm}

\noindent We describe several problems of interest to condensed matter physics
which could be solved by the massless form-factors approach, as perturbations
of integrable models. The basic unperturbed model is the massless sine-Gordon
model with the 
interaction  (cosine) term restricted to one point, which is integrable.
It is being perturbed by the second interaction term, $V(\phi)$, which destroys
the integrability:
\be
H={1\over 2}\int_0^\infty[\pi^2(x)+(\partial_x\phi)^2]dx + \mu\cos\beta\phi(0)
+\lambda V(\phi)
\ee
 Quasi-exact results can be obtained by making use of
the basis of massless quasiparticles of the sine-Gordon model. Namely,
one expands the evolution operator, $
S=\T e^{i\lambda\int V(\phi)}$,
in the various matrix elements of interest,
\be
\langle\T O_1O_2\dots\rangle=\langle\T O_1O_2\dots\rangle_0 +i\lambda\langle\T
O_1O_2\dots\int^t V(\phi)\rangle_0+\dots
\label{AVE}
\ee
to generate a perturbation series for the physical quantities of interest.
Further, for the evaluation of averages in (\ref{AVE}) one inserts the full
set of intermediate states -- massless integrable quasiparticles of the 
unperturbed sine-Gordon model:
\be
\langle O_1O_2\dots\int V(\phi)\rangle=\sum_{n,m,l\dots}\langle 0|O_1|n\rangle
\langle n|O_2|m\rangle\dots\langle l|\int^t V(\phi)|0\rangle
\ee
The form-factors $\langle n|O|m\rangle$ are known exactly.
While generating the perturbation series, the following normalization
conditions should be respected: 
\begin{eqnarray}
\langle 0|S|0\rangle &=& \langle 0|0\rangle \label{EQi} \\
\langle 1|S|1'\rangle &=& \langle 1|1'\rangle \label{EQii}
\end{eqnarray}
The first equation (\ref{EQi}) takes into account the renormalization
of the ground state energy (disconnected diagrams), 
while the second one (\ref{EQii})
controls the norm of one-particle states.  

\vspace{1cm}

1. {\it Quantum dot}

\vspace{5mm}

\noindent Matveev [Phys. Rev. B51 (1995) p.1743] argued that the physics of
Coulomb blockade in the quantum dots can be adequately described by
the following Hamiltonian, $H=H_0+H_C+H'$:
\be
H_0={1\over 2}\int^\infty_{-\infty}dx\left[
\pi^2(x)+ (\partial_xu)^2\right]
\label{HI}
\ee
\be
H_C=E_C[u(0)-N]^2
\label{HII}
\ee
\be
H'=-V\cos[2\pi u(0)]
\ee
where $E_C={e^2\over 2C_0}$ is the charging energy characterizing the dot,
parameter $N$ is proportional to the gate voltage which is adjustable.
Electron tunneling into the dot leads to the increase of energy by the
amount $E_C$. Therefore, the tunneling conductance is suppresed.
The gate voltage allows to control electrostatic energy of the dot:
\be
E_Q={(Q-eN)^2\over 2C_0}
\ee
where $Q$ is the total chatge of the dot. At the values of the
gate voltage corresponding to $N=n+1/2$ the energies of states with
charges $en$ and $e(n+1)$ are equal, leading to the favorable
conditions in the electron transport through the dot. Therefore, one
observes periodic resonanses of the conducatance as a function of the gate 
voltage.
The term $H'$ describes quantum tunneling (backscattering) effects, while
the term $H_C$ describes electrostatic charging effects. The Coulomb
blockade shows up in the oscillations of the ground state energy or the average
charge $\langle Q\rangle$ as a function of $N$, which can be calculated. 
Experimentally the
capacitance $C=\partial^2E/\partial N^2$ can be measured.    
At the perfect transmission, $V=0$, the ground state energy is $N$-independent,
as can be seen from (\ref{HI})-(\ref{HII}) by the change of variables
$u\to u+N$. The Coulomb blockade is
completely suppressed. On the contrary, for the perfect reflection, $V\to\infty$,
the dependence on $N$ is the most drastic, for the field is pinned at zero
by the value $n$ corresponding to one of the equivalent minima, 
and a departure of $N$ from integer $n$ leads to the 
increase of energy given by the term $H_C$,
thus lifting the degeneracy. At the values $N=n+1/2$ two
neighboring minima still remain degenerate, leading to the drastic 
changes in the
transport properties. 
 Matveev in his work calculates perturbative corrections
to the ground state energy. He treats the term $H'$ as a small perturbation
and uses the exact eigenstates of $H_0+H_C$. The form-factors of sine-Gordon
model allow, on the contrary, treat exactly the $H'$ term, starting
from the integrable model $H_0+H'$ and to perturb it by $H_C$ term.
The required form-factors of the field $u$, originating from the expression
for the ground state energy correction,
$\langle 0|\T\exp(i\int dt H_C)|0\rangle$, are available.

\vspace{1cm}

2. {\it Quantum Hall interferometer.}

\vspace{5mm}

\noindent Chamon et al [cond-mat/9607195] discussed the device based on the 
quantum
Hall bar with two constrictions which allows observation of interesting
interference effects. The underlying Hamiltonian is $H=H_0+H_{tun}$:
\be
H_0={1\over 8\pi}
\int^\infty_{-\infty}dx\left[
\pi^2(x)+ (\partial_x\phi)^2\right]
\label{HIa}
\ee
\be
H_{tun}=\gamma_1\cos\theta(0) +\gamma_2\cos[\theta(d)+\Phi_0]
\label{HT}
\ee
where the two fields $\theta, \phi$ are related to the left-right moving
fields $\phi_L, \phi_R$ as follows:
\be
\sqrt{2}\phi=\phi_L+\phi_R 
\ee
\be
\sqrt{2}\theta=\phi_L-\phi_R
\ee
The quantity of interest is the tunelling current $I_t(\Phi_0,d)$
as a function of flux $\Phi_0$ and distance $d$. For the current $I_t$
it is sufficient to take its linear response limit, given by Kubo's
formula. The current-current correlators can be obtained using
form-factors method and treating one of the cosine terms in (\ref{HT}) as
a perturbation of integrable sine-Gordon model.

\vspace{1cm}

3. {\it Double barrier in the Luttinger liquid.}

\vspace{5mm}

\noindent Kane and Fisher [Phys.Rev. B46  (1992) 15233] studied various
tunneling effects through the barriers in Luttinger liquid.
The double barrier action reads:
\begin{eqnarray}
S&=&S_0+V\int
d\tau\left\{
\cos[2\sqrt{\pi}\theta(\tau,x=0)]+\cos[2\sqrt{\pi}\theta(\tau,x=d)
+k_Fd]\right\} \nonumber \\
&+&{V_G\over\sqrt{\pi}}\int d\tau [\theta(\tau, x=d)-\theta(\tau, x=0)]
\end{eqnarray}
where $S_0$ is the action of the pure Luttinger liquid.
In the last term the total number of particles between two barriers is coupled
to the chemical potential of the island, $V_G$. In analogy with the problem
of quantum dot above, one can adjust $V_G$ to have the energy cost of adding
another particle vanish. Then, one expects a resonant transmission. The mass
scale $M=k_F/gd$ measures the fluctuations of the charge on the island:
\be
\langle n\rangle = {k_Fd\over 2\pi}+{V_G\over M}
\ee
The first term in the above formula is just the background density.
The resonanse is achieved when $\langle n\rangle=$half-integer.
In the strong barrier limit, $V\gg M$, the total charge of the island
prefers to be a particular integer times $e$, since
the discreteness of the electron charge is important.
The Coulomb blockade supresses the transport through the island.
When $\langle n\rangle$ is tuned to be a half-integer, an additional
symmetry is present which leads to the degenerate states of different charge.
In the opposite, weak barrier limit, $V\ll M$, following Kane and Fisher 
one can integrate out small fluctuations of the charge around the value
$\langle n\rangle$ and obtain an effective action:
\be
S=S_0+aV\int d\tau\cos[2\sqrt{\pi}\theta(0)] + {bV^2\over 2M}\int
d\tau\cos[4\sqrt{\pi}\theta(0)]
\label{eq}
\ee
where $a,b$ are dimensionless constants. The second cosine term in the last
formula is irrelevant for $g>1/4$, while the first one is relevant.
 At small temperatures the effective barrier strength $a$
grows, making the conductance to vanish at $T=0$. As for the effective coupling
$b$, it decreases, leading thus to the perfect conductance. Thus, by fine-tuning
the single parameter
$a=0$ (corresponding to the vanishing of the $2k_F$ Fourier component
of the double-barrier scattering potential $V(x)$),
one achieves the resonance. Qualitatively, the physical meaning of
the resonance is the following: forbidding the one-electron backscattering 
processes ($\hat{V}(2k_F)=0$), one creates favorable conditions
for the coherent transport of electron pairs ($\hat{V}(4k_F)\neq 0$).
It is interesting to find the conductance as a function of $a,b$ by the
form-factor method, treating one of the cosine terms in (\ref{eq}) 
as perturbation.

\vspace{1cm}

4. {\it Potential difficulties.}

\vspace{5mm}

\noindent In all the three problems one has to deal with IR divergences in the
form-factor expansion, since the perturbing terms are either
$\cos\phi$ or field $\phi$ itself. Such  divergences are peculiar
for the masssless theories. Lesage and Saleur [cond-mat/9608112]
have discussed already the situation. The partial resummation of 
the most divergent terms of the perturbation series is necessary.

In the first problem it is not clear that the perturbation theory
in $E_C$ will describe correctly the interesting regime of large $E_C$,
although the rapid convergence of the form-factors series leaves us
some hope that one is allowed to reach large enough values of $E_C$
while the terms of the order of $E^2_C$ and higher still can be 
neglected.

In the second problem the usual trick of changing the basis of fields
to the odd and even fields does not work with two constrictions.
The odd and even fields are not decoupled any longer.

In the third problem one has to deal with the irrelevant perturbation,
for which the form-factors are not quite known. One can in principle
overcome this
difficulty by an analytic continuation from the space of relevant perturbations.
The breather form-factors can be analytically continued from the sinh-Gordon
model, while kink and anti-kink form factors can be obtained by continuing
analytically the recent results of Lukyanov [hep-th/9703190].

\end{document}